# Antibonding and Electronic Instabilities in GdRu$_2$X$_2$ (X = Si, Ge, Sn): A New Pathway Toward Developing Centrosymmetric Skyrmion Materials

Dasuni N. Rathnaweera,[1] Xudong Huai,[1] K. Ramesh Kumar,[1] Sumanta Tewari,[2] Michał J. Winiarski,[3] Richard Dronskowski,[4] and Thao T. Tran[1,*]

[1]Department of Chemistry, Clemson University, Clemson, South Carolina, 29634, United States
[2]Department of Physics, Clemson University, Clemson, South Carolina, 29634, United States
[3]Faculty of Applied Physics and Mathematics and Advanced Materials Center, Gdansk University of Technology, ul. Narutowicza 11/12, 80-233 Gdansk, Poland
[4]Chair of Solid-State and Quantum Chemistry, Institute of Inorganic Chemistry, RWTH Aachen University, 52056 Aachen, Germany

**ABSTRACT:** Chemical bonding is key to unlocking the potential of magnetic materials for future information technology. Magnetic skyrmions are topologically protected nano-sized spin textures that can enable high-density low-power spin-based electronics. Despite increasing interest in the discovery of new skyrmion hosts and their characterization, the electronic origins of the skyrmion formation remain unknown. Here, we study GdRu$_2$X$_2$ (X = Si, Ge, Sn) as a model system to study the connection among chemical bonding, electronic instability, and the critical temperature and magnetic field at which skyrmions evolve. The nature of the electronic structure of GdRu$_2$X$_2$ is characterized by chemical bonding, Fermi surface analysis, and density of energy function. As X-*p* orbitals become more extended from Si-3*p* to Ge-4*p* and Sn-5*p*, improved interactions between the Gd spins and the [Ru$_2$X$_2$] conduction layer and increased destabilizing energy contributions are obtained. GdRu$_2$Si$_2$ possesses a Fermi surface nesting (FSN) vector [***Q*** = (*q*, 0, 0)], whereas GdRu$_2$Ge$_2$ displays two inequivalent FSN vectors [***Q*** = (*q*, 0, 0); ***Q**$_A$* = (*q*, *q*, 0)] and GdRu$_2$Sn$_2$ features multiple ***Q*** vectors. In addition, competing ferromagnetic and antiferromagnetic exchange interactions in the Gd plane become more pronounced as a function of X. These results reveal some correlation among the electronic instability, the competing interaction strength, and the temperature and magnetic field conditions at which the skyrmions emerge. This work demonstrates how chemical bonding and electronic structure enable a new framework for understanding and developing skyrmions under desired conditions that would otherwise be impossible.

**INTRODUCTION**

Understanding how chemical bonding and electronic instability influence magnetic phase transitions is vital for developing materials for various uses, such as spintronics, sensors, and quantum technologies.[1-10] Skyrmions are dynamic, particle-like magnetic states capable of twisting and turning in a unique way, with sizes ranging from a few nanometers to ~100 nm, and exhibit unique properties arising from their non-trivial topology. Their topologically protected properties provide a promising platform for studying the interaction between electronic effects and magnetic phase transitions while improving our understanding of novel states of matter.[11-18] The topological protection allows skyrmions to maintain their unique properties even in the presence of defects in real materials, offering new opportunities for developing next-generation information carriers and memory architectures.[19-21]

Skyrmions evolution in noncentrosymmetric magnets is driven by antisymmetric Dzyaloshinskii–Moriya interactions,[22-33] whereas in the case of centrosymmetric magnetic metals the formation is facilitated by a delicate balance between competing exchange interactions, the Ruderman-Kittel-Kasuya-Yosida (RKKY) interactions, frustrated exchange coupling, and dipolar interactions. The long-range RKKY exchange interaction $J(r) \sim \sin(2k_F r) / r^3$, where $k_F$ is the Fermi wavevector of conduction electrons, and $r$ is the distance between the magnetic moments, assisted by intra-orbital magnetic frustration, promotes skyrmion formation with small sizes (a few nm).[13,34-39]

Gd-based centrosymmetric tetragonal lattice systems, $GdRu_2X_2$ ($Gd^{3+}$, $S = 7/2$, $L = 0$, X = Si and Ge), have been demonstrated as skyrmion hosts with rich magnetic phase diagrams.[40-44] This is in part attributable to the unique crystal structure and chemical flexibility, allowing an array of atomic substitutions (**Figure 1a**). Recent theoretical studies on skyrmion hosts adopting centrosymmetric tetragonal lattices suggested several microscopic origins for skyrmion emergence, such as RKKY exchange and the interplay of four-spin interactions associated with $s$–$d$ or $s$–$f$ coupling mediated by itinerant electrons.[42] In addition, some other theoretical works showed interorbital frustration originating in Gd-$d$-$f$ spins can stabilize a spin structure with a finite modulation vector, $Q$.[45] Magnetic frustration, anisotropy, high-order spin interactions and Fermi surface nesting (FSN) have been recognized as essential physical parameters to induce phase transitions to skyrmions at a given field and temperature.[46,47] The elegant studies provided valuable insights into the microscopic origins of centrosymmetric skyrmion materials; however, chemical connections to critical temperature and magnetic field conditions at which skyrmions emerge remain elusive. This significantly hinders the materials development of skyrmions that may evolve at room temperature and zero field.

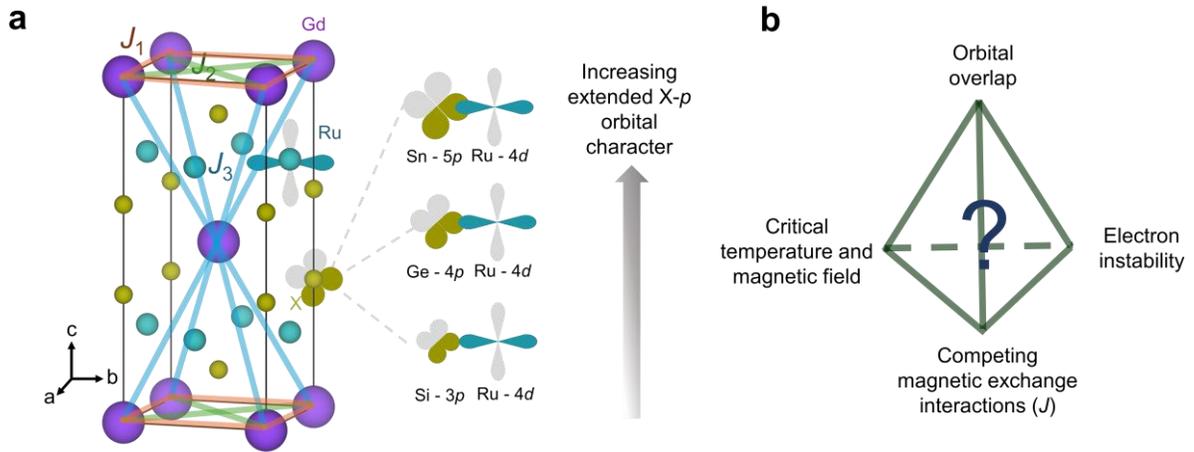

**Figure 1.** (**a**) Crystal structure of GdRu$_2$X$_2$ (X = Si, Ge, and Sn) showing exchange interactions and increased dispersion of X-$p$ orbitals going from Si-3$p$ to Ge-4$p$ and Sn-5$p$, and (**b**) Potential connections among orbital overlap, electron instability, and competing exchange interactions, and critical temperature and magnetic field conditions at which skyrmions emerge.

It has been demonstrated that in GdRu$_2$Si$_2$, indirect RKKY interactions stabilize equivalent magnetic modulation vectors, resulting in a square skyrmion lattice at 2 T ≤ $\mu_0H$ ≤ 2.5 T, 2 K ≤ $T$ ≤ 20 K, with the smallest diameter of 1.9 nm among the known skyrmion materials.[42,48,49] Magnetic torque and resistivity measurements showed that FSN enhances the strength of the RKKY interaction, leading to a helical modulation ***Q*** = (0.22$q$, 0, 0) in GdRu$_2$Si$_2$.[50] Angle-resolved photoelectron spectroscopy measurements in conjunction with density functional theory calculations revealed a nested Fermi surface (FS) band at the corner of the Brillouin zone that is responsible for the skyrmion formation in the Si material.[51] GdRu$_2$Ge$_2$—an isostructural compound—has recently been realized to feature the successive formation of two distinct skyrmion pockets.[40,41] Resonant X-ray scattering and magnetotransport studies suggested the presence of competing RKKY exchange interactions at inequivalent wavevectors, [***Q****$_A$* =($q$, 0, 0) and ***Q****$_B$* = ($q/2$, $q/2$, 0)] that drive such rich topological phase formation.[40] Our previous studies have demonstrated the evolution of two skyrmion pockets in GdRu$_2$Ge$_2$ at 2 K ≤ $T$ ≤ 30 K, 0.9 T ≤ $\mu_0H$ ≤ 1.2 T and 1.3 T ≤ $\mu_0H$ ≤ 1.7 T. It is worth noting that the skyrmions in GdRu$_2$Ge$_2$ form at higher temperatures and lower fields than that in GdRu$_2$Si$_2$. Electronic structure and exchange interaction evaluations revealed that the more extended Ge-4$p$ orbitals, compared to Si-3$p$, enhance competing exchange interactions, thereby making the phase transition to skyrmions more accessible in GdRu$_2$Ge$_2$ (at higher temperatures and lower fields).[41]

This work provides a systematic investigation into how electronic structure and chemical bonding manifest the temperature and magnetic field conditions for skyrmion evolution in the isostructural model system GdRu$_2$X$_2$ (X = Si, Ge, and Sn). While the Sn compound has not been experimentally realized due to its thermodynamically unfavorable formation energy (**Figure S1**), its hypothesized crystal structure is optimized for this study using variable cell calculations. Our research goal is to provide some answers to the central scientific questions: *What* underlying chemical and electronic parameters influence the formation of skyrmions in GdRu$_2$X$_2$? *How* do they manifest in critical temperature and magnetic field conditions at which skyrmions emerge? (**Figure 1b**). In this study, we employ density functional theory (DFT) calculations to investigate chemical bonding via crystal orbital Hamilton population (COHP) and crystal orbital bond index (COBI), analyze Fermi surface nesting, probe electronic instability through density of energy (DOE), and use total-energy methods to elucidate the exchange interactions in the model system. This approach enables us to delve into the impact of Si-3$p$/Ge-4$p$/Sn-5$p$ on the orbital overlap and electronic structure of GdRu$_2$X$_2$ and connect the chemical bonding concepts to skyrmions formation.

**RESULTS AND DISCUSSION**

We have previously studied the skyrmion host GdRu$_2$Ge$_2$, which adopts ThCr$_2$Si$_2$-type structure—a centrosymmetric tetragonal space group *I4/*mmm.[41] For the magnetoentropy mapping, electrical transport and heat capacity data of the Ge material, the reader is invited to visit the reference.[41] The previous results lay some groundwork for our systematic studies on the isostructural centrosymmetric magnets GdRu$_2$X$_2$ (X = Si, Ge, and Sn), of which the structure includes the Gd square lattice connected to [Ru$_2$X$_2$] layers (**Figure 1a**).

To get insight into how the electronic structure of GdRu$_2$X$_2$ determines its physical properties, pseudo-potential spin-polarized DFT calculations were performed using the Quantum Espresso software package.[52] The crystal structure information of GdRu$_2$X$_2$ used in the DFT computations is provided in **Table S1**. The band structure and density of states (DOS) results clearly demonstrate some common electronic features in GdRu$_2$X$_2$ (**Figure 2**). The spins of the Gd-4$f$ states are polarized, which then polarizes the Ru-4$d$ and X-$p$ states. The contribution of Gd-4$f$ states is localized, deep in low energy ~ -8 eV for majority spins, and slightly above the Fermi level ($E_F$) energy ~ 4 eV for minority spins. The spin-polarized band structure and density of states of GdRu$_2$X$_2$ display a metallic behavior, where multiple bands cross $E_F$ and finite DOS at $E_F$. These features prove that interactions between the localized Gd-4$f$ magnetic moments are mediated through the itinerant electrons—RKKY interactions. Around $E_F$, the bands mostly comprise the Gd-5$d$, Ru-4$d$, and X-$p$ states. The band structures exhibit an increase in overall band dispersion going from Si-3$p$, Ge-4$p$, and Sn-5*p*. As a result, the extended X-*p* orbital character generates more diffused features in Gd-*d* and Ru-*d* orbitals, improving the interaction between Gd-X and Ru-X (**Figure 2**).

DOS analysis is helpful in describing the atom projected state contribution; however, it does not capture the phase relationships among the orbitals involved in the overlapping wavefunctions. To extract the information on bonding characters (bonding = constructive interference of wavefunctions, nonbonding = zero interference, and antibonding = destructive interference) and how the microscopic mechanisms can influence skyrmion formation, we used the Local Orbital Basis Suite Towards Electronic-Structure Reconstruction (LOBSTER) program to reconstruct the local-orbital picture from projector-augmented wave (PAW) wavefunctions.[53-57]

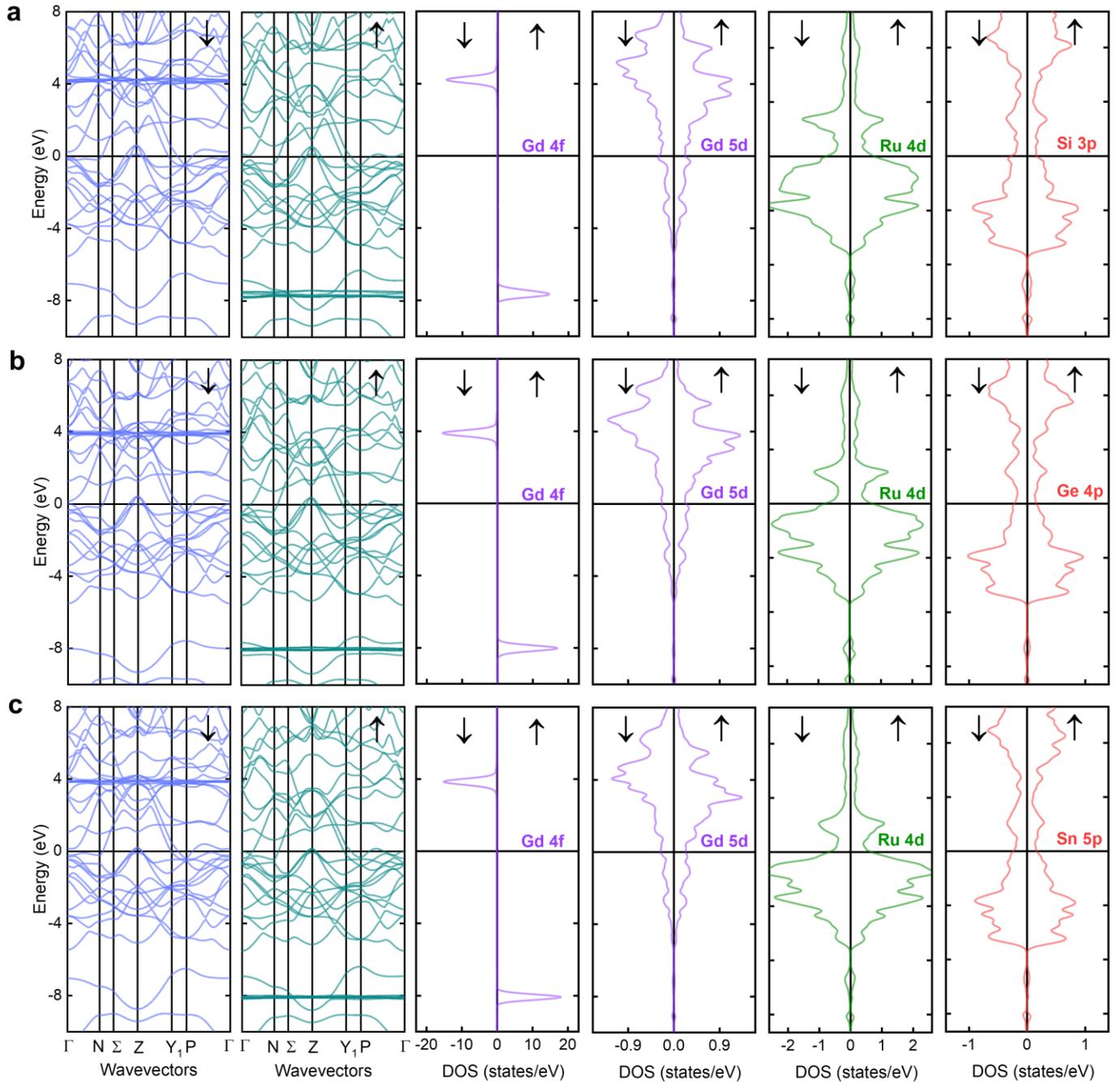

**Figure 2**. Spin-polarized band structures showing bands around the Fermi level and spin-polarized DOS of (**a**) GdRu$_2$Si$_2$, (**b**) GdRu$_2$Ge$_2$, and (**c**) GdRu$_2$Sn$_2$.

The projected COHP (pCOHP) curves (**Figure 3**) indicate bonding character (-pCOHP > 0) for Gd–Gd, Gd–X, and X–X bonds and antibonding character (-pCOHP <0) for Ru–Ru and Ru–X bonds around $E_F$. The integrated COHP (ICOHP) value of Gd–Gd decreases going from the Si to Ge and Sn material, suggesting a reduced overlap of the Gd orbitals. For the [Ru$_2$X$_2$] conduction layer, the increased ICOHP value of Ru–X implies an improved overlap of the Ru-$d$ and X-$p$ orbitals as X changes from Si to Ge and Sn. These combined features give some hints about improved interaction between the Gd-4$f$ localized electrons and the [Ru$_2$X$_2$] layer from Si to Ge and Sn.

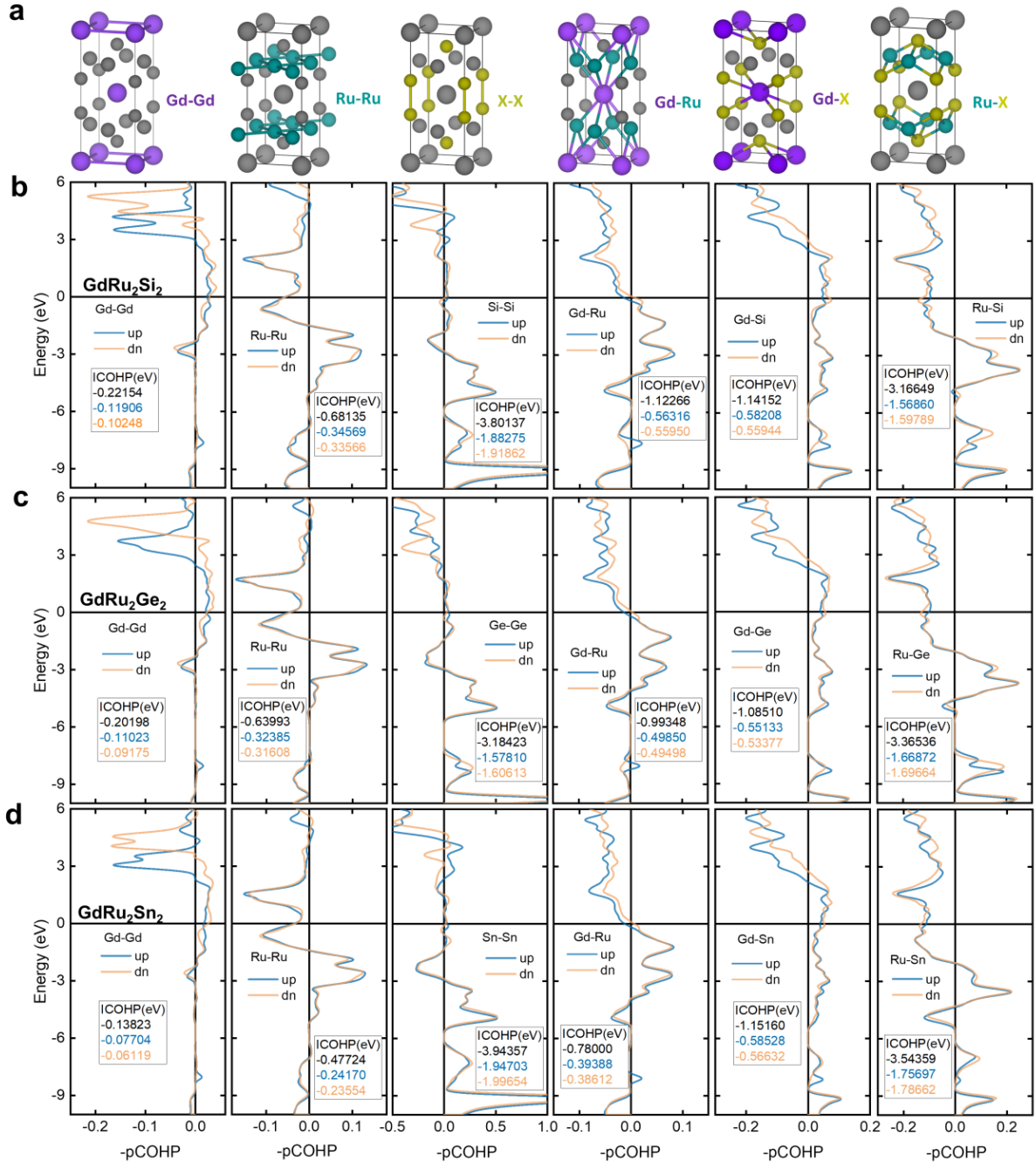

**Figure 3.** Projected crystal orbital Hamilton population (-pCOHP) curves for (**a**) relevant bonding environments, (**b**) GdRu$_2$Si$_2$, (**c**) GdRu$_2$Ge$_2$, and (**d**) GdRu$_2$Sn$_2$ with their integrated values (ICOHP).

To quantitatively analyze the quantum-chemical bond order in solids, integrated COBI (ICOBI)[58,59] were calculated for all bonds (**Figure S3**). The Ru–X and X–X bonds have the greatest ICOBI values of ~0.5, about a half bond, indicating a strong covalent bonding character. The small ICOBI value of Gd-Gd (~0.03) implies negligible pairwise interactions. Overall, the COBI results are in line with and complementary to COHP.

We constructed molecular orbital (MO) diagrams for GdRu$_2$X$_2$ and [Ru$_2$X$_2$] (**Figure 4a-b**) with LOBSTER.[60] Symmetries of the MOs around $E_F$ are similar (nondegenerate a' and a"). Nevertheless, the MO diagrams become more diffuse as X goes from Si to Ge and Sn, lowering the energy levels of the MOs above $E_F$. It is worth noting that the MOs around $E_F$ are mainly derived from the [Ru$_2$X$_2$] layers. **Figure 4c-d** depicts a sketch for bonding characters between Ru-$d$ and X-$p$ orbitals in [Ru$_2$X$_2$]: antibonding at **Γ** and bonding at **P**—a result deduced from the combination of the MOs and the band structure of GdRu$_2$X$_2$. The tunability of the MO energy levels and the antibonding character at **Γ** imply an ability to modulate phase

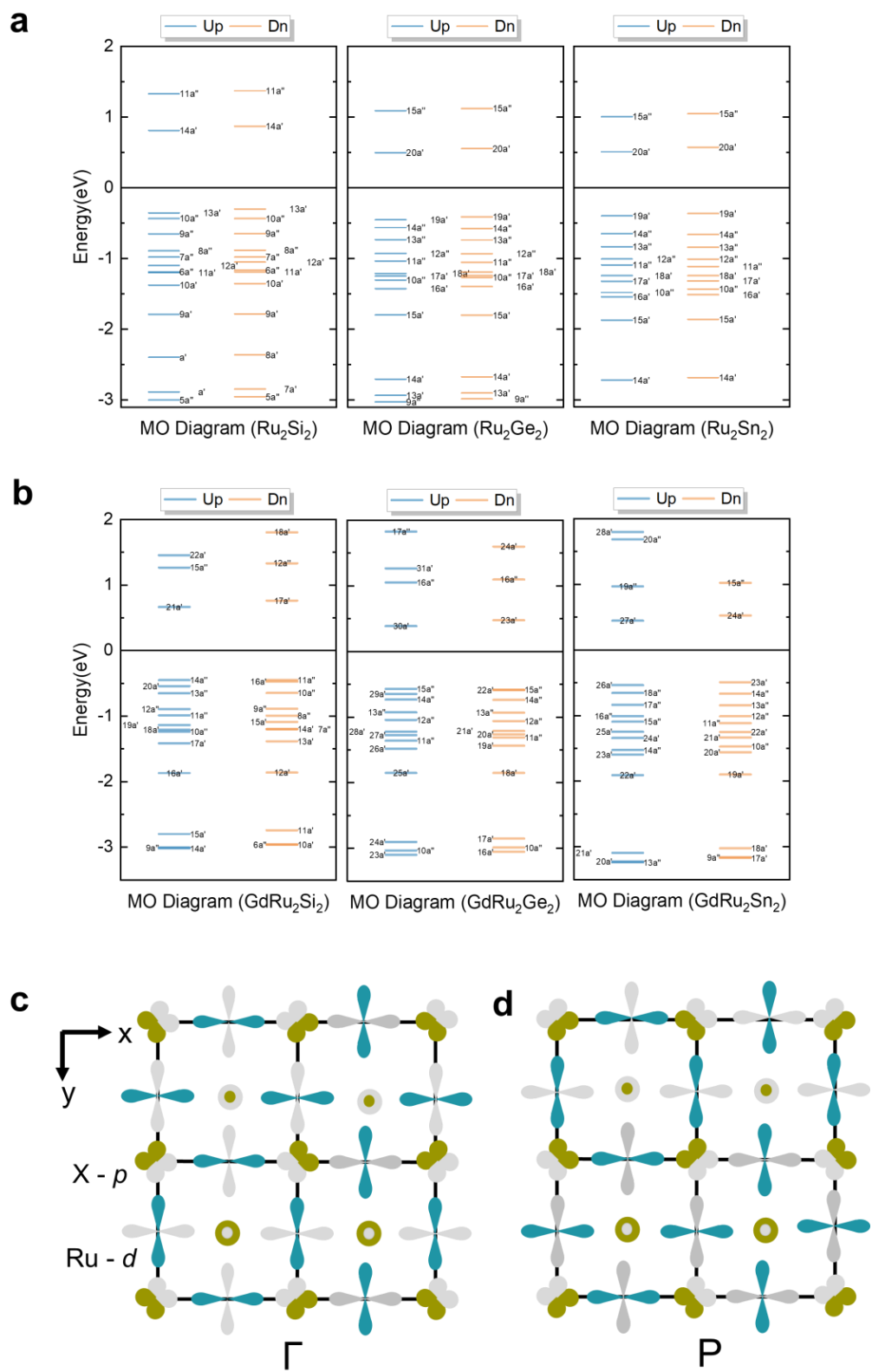

**Figure 4.** (**a**) Molecular orbital diagrams for GdRu$_2$X$_2$. (**b**) Molecular orbital diagrams for [Ru$_2$X$_2$]. (**c**) Orbital overlap in the [Ru$_2$X$_2$] layer showing antibonding at **Γ** and (**d**) bonding at **P**.

transition conditions for GdRu$_2$X$_2$.

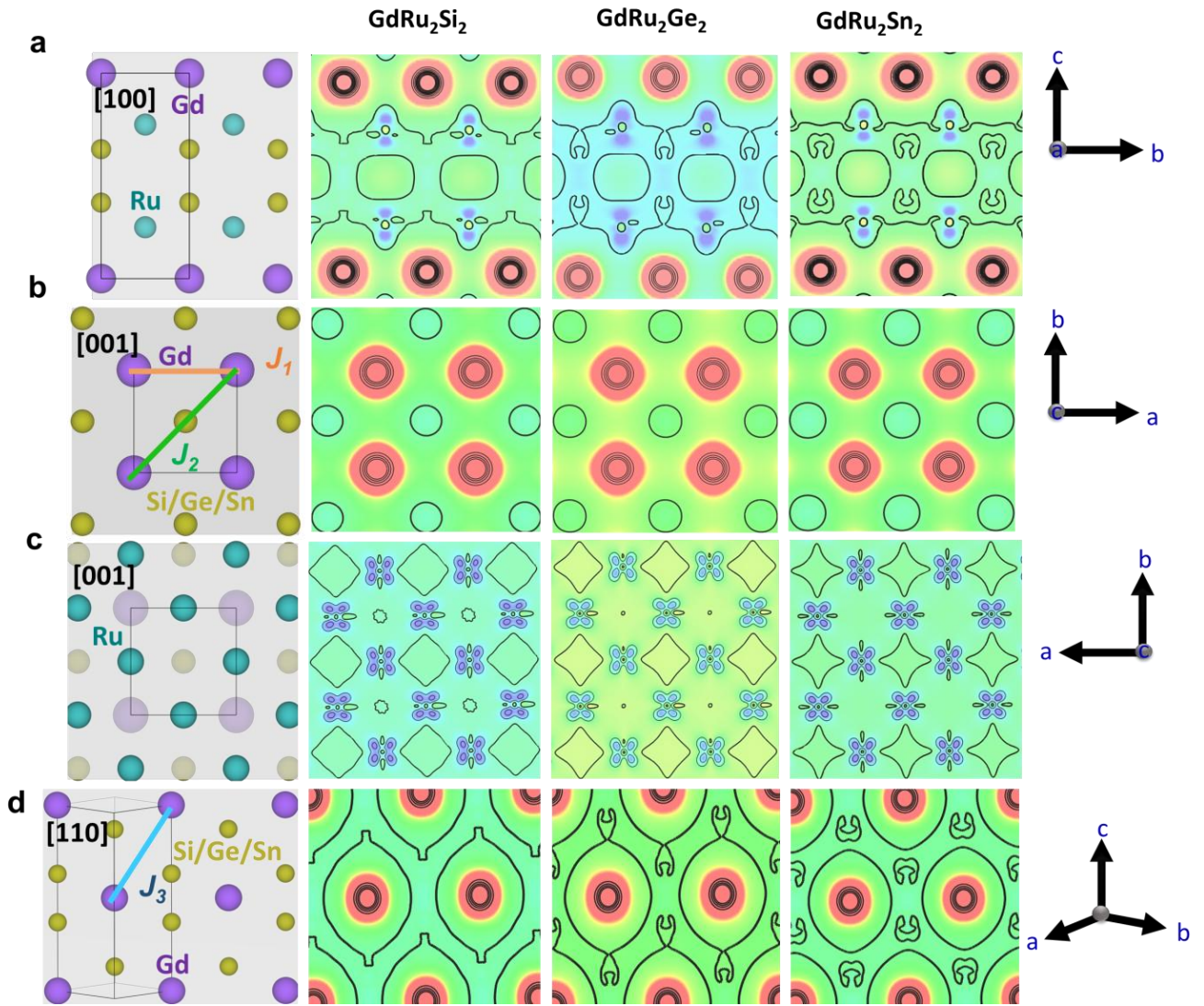

**Figure 5**. Spin density map of GdRu$_2$X$_2$ (**a**) [100] plane cutting through Gd, Ru, and X atoms, (**b**) [001] plane cutting though Gd and X atoms, (**c**) [001] plane cutting in between Ru and X atoms, and (**d**) [110] plane cutting through Gd atoms (X= Si, Ge, and Sn).

Spin, charge, and orbital are the three fundamental components of an electron. Next, we examine spin and charge density maps while considering orbital contribution. Spin density maps ($\rho_{up}$ - $\rho_{down}$) projected on selected lattice planes illustrate supplementary evidence for the effect of the X-$p$ orbitals on spin polarization (**Figure 5**).[61] In the [100] plane cutting through Gd, Ru, and X (**Figure 5a**), the spin polarization of the [Ru$_2$X$_2$] layer is enhanced as X changes from Si to Ge and Sn. **Figure 5b-c** shows the [001] plane slicing through Gd and X and between Ru and X, respectively. The spin density of Gd and Ru in these projections becomes more polarized, following the trend GdRu$_2$Si$_2$ < GdRu$_2$Ge$_2$ < GdRu$_2$Sn$_2$. The tunable spin polarization as a function of X-3$p$/4$p$/5$p$ orbitals suggests the feasibility of modifying interactions between the Gd localized magnetic spins and the [Ru$_2$X$_2$] conduction layer—ingredients for RKKY interactions. As shown in **Figure 5a-d**, the Gd site exhibits a strong, red-colored, spherically symmetric contour pattern, indicating well-localized Gd-4$f$ magnetic moments. In the vicinity of the Gd moment, the Ru–X layer exhibits spin density in the blue-purple color, reflecting strong hybridization between the Ru-4$d$ states and X-$p$ states (**Figure 5a**). If we interpret the blue-purple region as induced spin polarization due to the RKKY interaction, it becomes evident from **Figure 5a** that the substitution of the X atom from Si to Ge to Sn leads to subtle changes in both the intensity and the spin density contours along the [100] direction. The contour lines emerging along the [100] projection indicate spin density elongation and are associated with the FSN vector $Q$ = ($q$,0,0) experimentally observed for GdRu$_2$Si$_2$ and GdRu$_2$Ge$_2$.[40,42] Additionally, short, wave-like modulations in the form of spin density lobes along the $b$-axis are observed in **Figure 5a**. The spin density wave along the $a$-axis ($a$ and $b$ are equivalent in tetragonal systems) can be linked to the experimentally observed propagation vector $Q$ = (0.22, 0, 0). Further, the Ge variant shows visibly stronger polarization and a tendency to diffuse along the $c$-axis, suggesting that enhanced orbital spatial extension of Ge-4$p$ and Sn-5$p$ facilitates increased hybridization and thereby supports FSN. Another interesting feature is observed along the [110] direction in the spin density contours. For the Si system, we observe nearly isotropic spin polarization at the Gd site. In contrast, the Ge and Sn variants exhibit reconfigured spin density contours around Gd, indicating the presence of additional spin density modulations (**Figure 5d**). This observation aligns with experimental reports of inequivalent multiple modulation vectors in GdRu$_2$Ge$_2$. Since we do not observe a clear antiferromagnetic-like coupling between the 4$f$ moments and the Ru–X layer as seen in the [100] modulation, the spin density contours along [110] may originate from inter-orbital frustration between Gd-4$f$ and Gd-5$d$. Similar spin density features are observed in the [110] projection (**Figure S5**), connecting to the $Q_A$ = ($q$, $q$, 0) FSN vector in GdRu$_2$X$_2$. A more detailed description of FSN is discussed in the subsequent sections.

In addition to the spin density, charge density maps visually represent the distribution of electronic charge within the GdRu$_2$X$_2$ crystal (**Figure 6**).[61] **Figure 6a** highlights the [110] projection on the Gd and X planes. As X changes from Si to Ge to Sn, the charge density of the X–X dumbbells increases, enhancing the charge anisotropy of Gd. Meanwhile, **Figure 6b** shows the [110] plane cutting through the Ru layer, where the charge distribution of Gd appears more diffused. **Figures 6c** and **6d** show the [001] projections on the Gd and X plane and the plane between Ru and X, respectively. As X goes from Si to Ge and Sn, the charge density of Gd and [Ru₂X₂] increases and becomes more dispersed. This indicates improved bonding interactions between Gd and [Ru₂X₂].

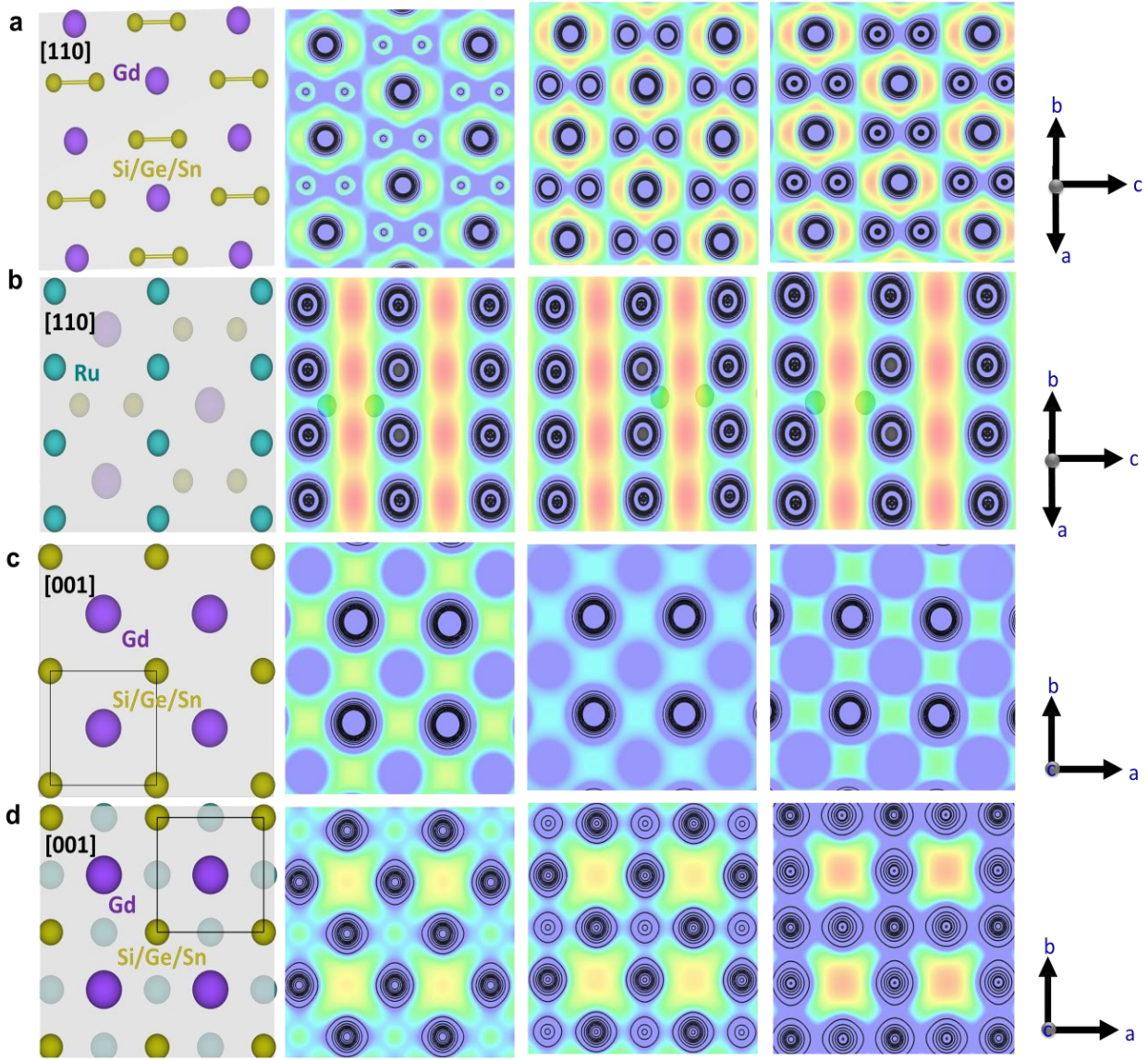

**Figure 6**. Charge density map of GdRu$_2$X$_2$ (**a**) [110] plane cutting through Gd and X atoms, (**b**) [110] plane cutting through Ru layer, (**c**) [001] cutting through Gd and X atoms, and (**d**) [001] plane in between Ru and X atoms. (X= Si, Ge, and Sn).

Electronic instability around $E_F$ in metals gives rise to phase transitions to novel states of matter under external perturbations, such as temperature and field.[62,63] To describe how itinerant electrons behave under the influence of various interactions in GdRu$_2$X$_2$, we extracted the Fermi surface from the bands that cross $E_F$. The Fermi surface topology and potential nesting in the Si and Ge materials and Gd$_2$PdSi$_3$ have been realized to enable spin spiral magnetic ordering and double-***Q*** density wave modulation.[40,46,51,64] There are situations where sections of the Fermi surface are parallel (or nearly parallel) and can be connected by a single nesting vector $Q$, leading to instabilities such as charge and spin density waves in materials. There is a general agreement in the literature regarding the FSN driving the helical and skyrmion phases. However, minor discrepancies arise regarding the specific nesting vector. Despite several studies, the precise modulation vector and the specific portion of the FS responsible for nesting remain unresolved in GdRu$_2$X$_2$ (X = Si, Ge).

To further understand FSN in GdRu$_2$X$_2$ (X = Si, Ge, Sn) and whether and how the modification of X-$p$ orbitals influences nesting vectors, we examine the FS and the Lindhard response function (LRF) for GdRu$_2$X$_2$. **Figures 7** and **8** present FS diagrams for both spin-up and spin-down channels and corresponding LRF projected along [100] and [111], respectively. Our calculations showed five distinct FSs in the Si and Ge and only four FSs in the Sn variant (**Figure S6**), agreeing well with the band structures. The presence of small, ellipsoidal FSs at the **Z** point is common in GdRu$_2$X$_2$ (**Figure S6**). These smaller ellipsoid pockets are identified as localized hole pockets and show parallel features at the **Z** point. However, LRF and other experimental evidence are absent for such modulation. A larger barrel-like FS (band 4 for Si/Ge and band 3 for Sn) is observed in all three compounds. This surface extends to the zone boundary and shows strong dispersion along the $k_x$ and $k_y$ directions with limited dispersion along the $k_z$ direction. **(Figure S6)**. These features are observed for both spin-up and spin-down channels. Additionally, band 5 (band 4 for Sn) shows a distinct difference between spin-up and spin-down channels. We observed a less dispersive parallel surface centered at the **X** point in the spin-down channel. In contrast, the spin-up band is well extended, (tubular type FS) within the Brillouin zone, suggesting significant band interaction, possibly due to the influence of 4$f$ and 5$d$ orbital mixing and/or multi-band effects of Ru-4$d$ (**Figure 7a, c and e**). This band primarily arises from Gd-5$d$ and Ru-4$d$ states, crosses $E_F$ at multiple $k$-points and displays linear dispersion just below $E_F$ and a flat band segment along the **Z-Y$_1$-P-Γ** direction, indicating potential topological features (**Figure 2**). Overall, this analysis highlights the similarities and differences in the FS characteristics in GdRu$_2$X$_2$. However, these characteristics alone cannot definitively determine the presence or absence of FSN in this system. Therefore, we employed the FermiSurfer software to compute the LRF from the conventional Fermi surface.

FSN and associated divergence in electronic susceptibility owing to lattice instability, spin density wave modulation or Friedel oscillations are understood in several low-dimensional systems.[62,65-67] Such electronic instabilities show up as a portion of parallel segments in the FS. However, these apparently parallel sections themselves are insufficient to draw conclusive evidence for FSN, as certain geometric and energy conditions are required to obtain FSN. One way to reliably determine a possible FSN is by calculating the Lindhard susceptibility function $\chi'(q)$ that describes the stability of the electron system. The Lindhard susceptibility function can be written as,

$$\chi'^{(q)} = \sum_k \frac{f(\varepsilon_k) - f(\varepsilon_{k+q})}{\varepsilon_k - \varepsilon_{k+q}} \quad (1)$$

where a $f(\varepsilon_k)$ and $f(\varepsilon_{k+q})$ are Fermi-Dirac probability distribution functions related to the occupancy of the states with energy $\varepsilon_k$ and $\varepsilon_{k+q}$, respectively.[67]

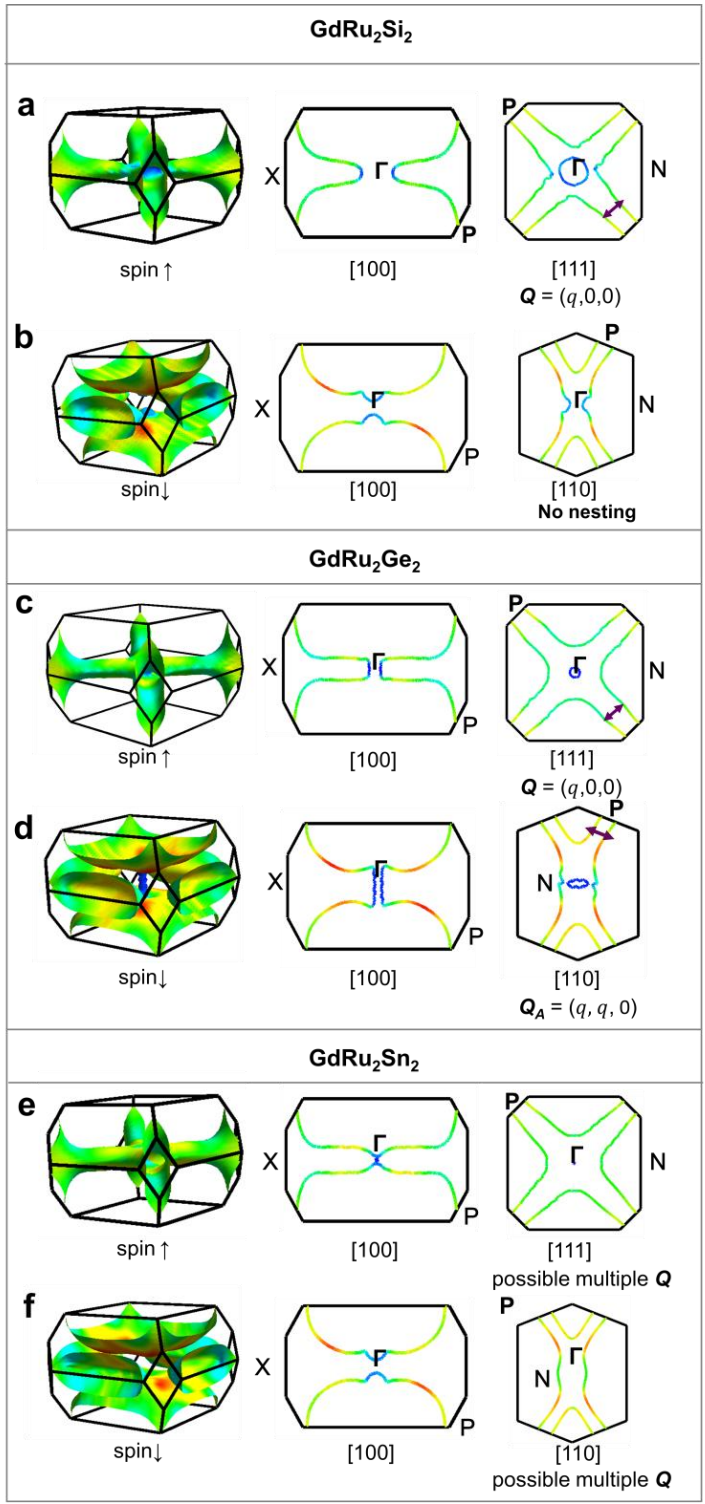

**Figure 7.** Fermi surface nesting with respective projections indicating nesting vector for (**a-b**) GdRu$_2$Si$_2$, (**c-d**) GdRu$_2$Ge$_2$, and (**e-f**) GdRu$_2$Sn$_2$, respectively.

The observed parallel regions centered around the **X** point, perpendicular to the **Γ-X** direction, suggest a nesting vector $Q$ = ($q$, 0, 0), (**Figure 7a** and **7c**) in excellent agreement with previous reports.[40,51] There are diffuse peaks in the Lindhard function at the **X** point

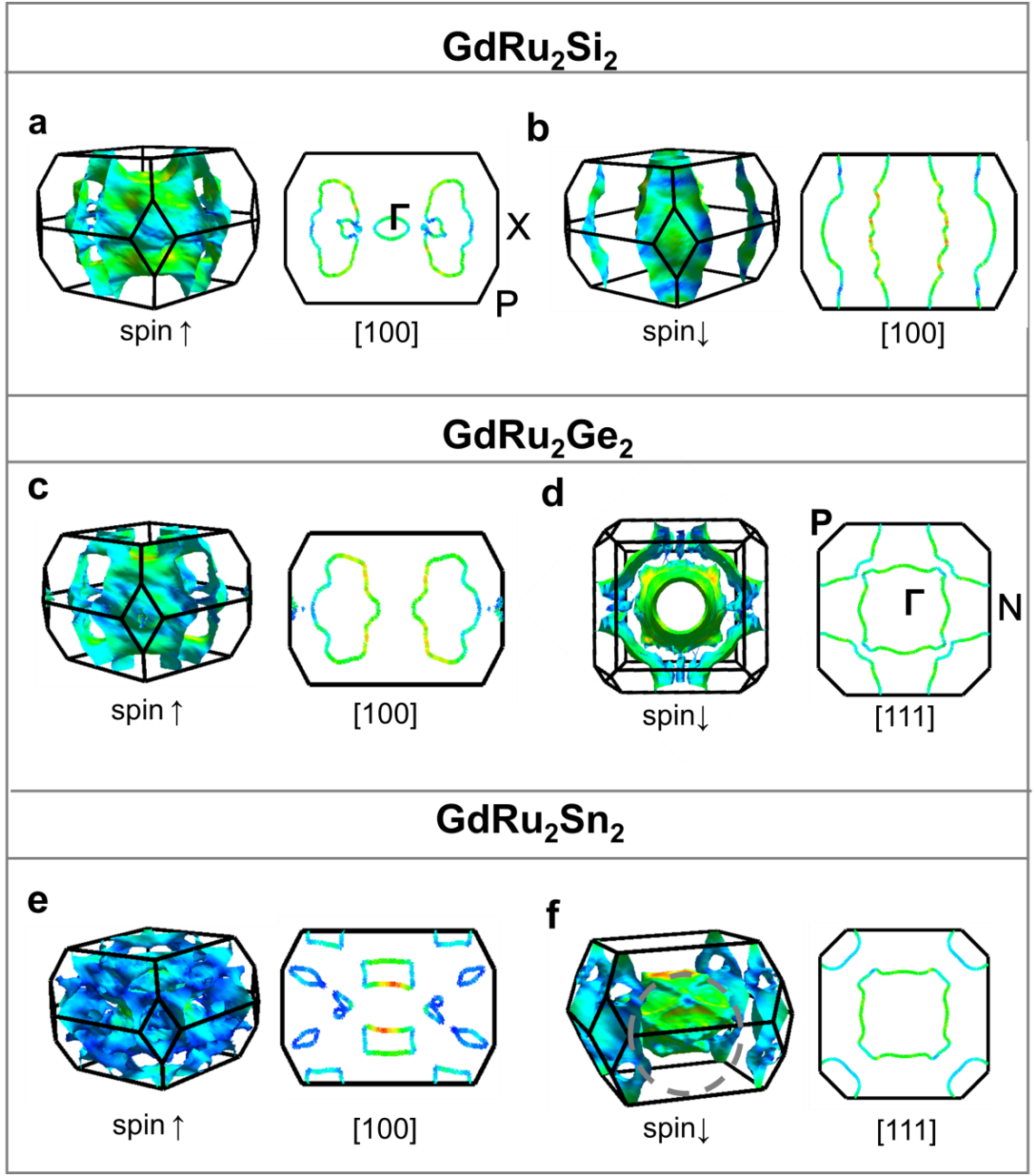

**Figure 8.** Lindhard response function with respective projections for (**a-b**) GdRu$_2$Si$_2$ (**c-d**) GdRu$_2$Ge$_2$, and (**e-f**) GdRu$_2$Sn$_2$ respectively.

which is consistent with the observation of the parallel surfaces for both spin-up and spin-down channels for all three compounds. The corresponding FSN vector is calculated to be $Q$

= ($q$, 0, 0) for both GdRu$_2$Si$_2$ and GdRu$_2$Ge$_2$ (**Figures 7a** and **7c**). In addition, Lindhard susceptibility peaks near the **N** point show FSN vectors of $\boldsymbol{Q_A}$ = ($q$, $q$, 0) and $\boldsymbol{Q_A'}$ = (-$q$, -$q$, 0) for GdRu$_2$Ge$_2$ (**Figure S7**). This multi-$Q$ Fermi surface nesting in GdRu$_2$Ge$_2$ is supported by the work by Yoshimoshi et al., which suggests that the competition of RKKY exchange interactions at inequivalent wavevectors drives the formation of such a rich topological phase.[40] In GdRu$_2$Ge$_2$ and GdRu$_2$Sn$_2$, there are parallel features along [101] and [-10-1] ($\boldsymbol{Q_B}$ = ($q$, $q$, 0) and $\boldsymbol{Q_B'}$ = (-$q$, -$q$, 0)), which is surprising as the dispersion along $k_z$ direction is minimal (**Figure S7**). This feature is observed only with Ge and Sn variants, whereas Si shows no nesting along these directions. This divergence may be connected to the formation of an additional skyrmion pocket observed only in Ge, not in Si. These two sets of inequivalent modulation vectors may interact with each other and can give rise to non-trivial magnetic interactions facilitating the skyrmion formation.

For GdRu$_2$Sn$_2$, the Lindhard susceptibility possesses diffuse features for spin-up and doubly degenerate susceptibility peaks and rhombohedral ridge patterns for spin-down (**Figure 8f**). These rhombohedral features are also observed in the spin-down channel for GdRu$_2$Ge$_2$. These complex features in the Lindhard response indicate multiple nesting wavevectors in GdRu$_2$Sn$_2$. This is attributable to increased interactions between the Gd-$f/d$ spins and the [Ru$_2$Sn$_2$] layer, as demonstrated by chemical bonding analysis. Since Gd is half-filled, any tendency to populate spin-down channels can create inter-orbital frustration, and such magnetic frustration, seen by the polarized conduction electron, can lead to magnetic modulation with a finite wave vector. Our FSN and LRF analysis capture the interplay between the RKKY interaction and conduction electron density modulation essential for the stabilization of topologically nontrivial spin states in GdRu$_2$X$_2$.

The observation of FSN validated by the LRF confirms the presence of electron instability in GdRu$_2$X$_2$. This argument is further examined by calculating the DOE. The DOE accounts for both interatomic and on-site atomic energy contributions as it integrates the entire electronic band structure with respect to energy (Eq (2)), unlike COHP, which focuses only on interatomic (pair-wise) interactions.[68] **Figure 9** shows the DOEs and the corresponding integrals. Overall, there are strongly destabilizing energy contributions (negative -DOE values), consistent with the electron instability underpinning FSN. The integrated DOE-band energy ($E_{\text{band}}$) for GdRu$_2$X$_2$ reveals that the chemical substitution of Si by Ge and Sn, separately, increases the overall destabilizing energy (**Figure 9**), as suggested by the Ru-Ru and Ru-X COHP curves (**Figure 3**). The $E_{\text{band}}$ curves of GdRu$_2$X$_2$ display similar shapes to one another, confirming their resemblance in chemical and crystal structures.

$$E_{\text{band}} = \int_{-\infty}^{E_F} \text{DOE}(E) dE \qquad (2)$$

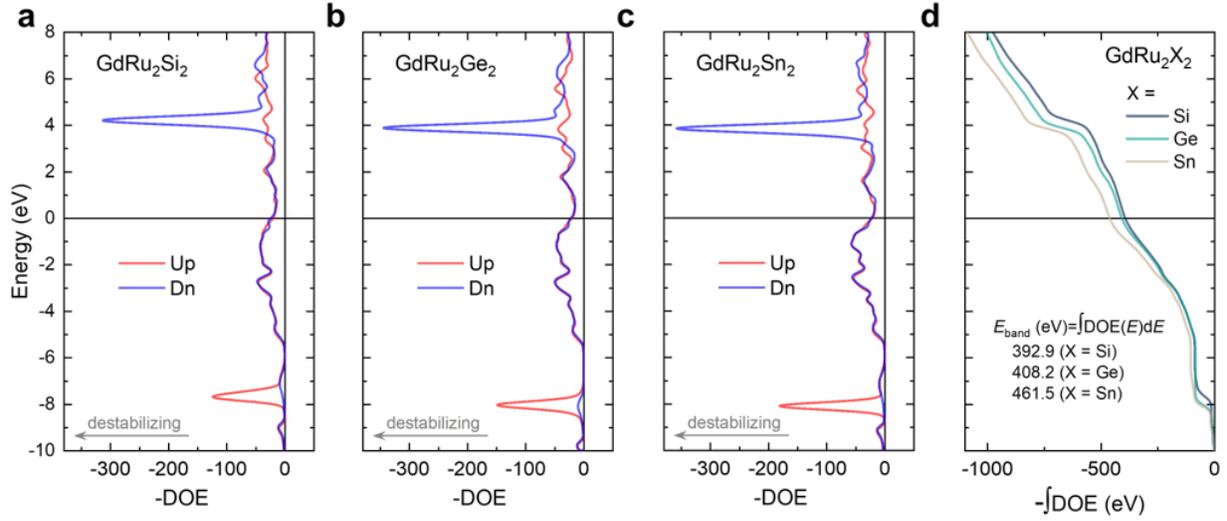

**Figure 9**. Density of energy analysis of (**a**) GdRu$_2$Si$_2$, (**b**) GdRu$_2$Ge$_2$, and (**c**) GdRu$_2$Sn$_2$. (**d**) Comparison of band energy ($E_{band}$) for GdRu$_2$X$_2$ where X = Si, Ge and Sn).

In addition to the bonding and DOE analysis, we study the exchange interactions in GdRu$_2$X$_2$. Building upon our earlier findings, [41] here we performed an energy-mapping analysis, using DFT calculations, as illustrated in **Figure 10**.[52,69-71] The Gd spins form two square sublattices within a unit cell. Our model incorporates the nearest-neighbor exchange interaction $J_1$ along the *a*- or *b*-axis, the exchange interaction $J_2$ within the *ab*-plane along the [110] direction, and the interaction $J_3$ between the Gd square sublattices. To accurately simulate an extended solid and avoid artificial interactions between periodic images of atoms, we used a (2*a*, 2*b*, *c*) supercell (containing four formula units) and six different spin-ordered states (**Figure 10**). The total spin exchange energy per supercell can be expressed by the following equations:[5]

$$E_1 = E_0 + (-8J_1 + 0J_2 + 0J_3) \cdot S^2 \qquad (3)$$
$$E_2 = E_0 + (0J_1 - 16J_2 + 0J_3) \cdot S^2$$
$$E_3 = E_0 + (0J_1 + 0J_2 - 8J_3) \cdot S^2$$
$$E_4 = E_0 + (8J_1 - 8J_2 + 0J_3) \cdot S^2$$
$$E_5 = E_0 + (8J_1 + 0J_2 + 0J_3) \cdot S^2$$
$$E_6 = E_0 + (16J_1 - 16J_2 + 0J_3) \cdot S^2$$

where the $E_0$ corresponds to the non-magnetic contribution to the total energy, and $S = 7/2$, the spin for Gd$^{3+}$. From these energies, the exchange interactions per four formula units can be calculated as:

$$J_1 = \frac{(E_6 - E_2)}{16S^2} \qquad (4)$$
$$J_2 = \frac{(E_5 - E_4)}{8S^2}$$
$$J_3 = \frac{(E_1 + E_5 - 2E_3)}{16S^2}$$

The *J*-coupling interaction can then be obtained:

**Table 1**. Calculated *J*-coupling constants and an inverse of the Mean Absolute Deviation (MAD$^{-1}$) for GdRu$_2$X$_2$ (X = Si, Ge, Sn)

| X Site | $J_1$ (K) | $J_2$ (K) | $J_3$ (K) | MAD$^{-1}$ |
|---|---|---|---|---|
| Si | -0.7(2) | -153.5(2) | -0.5(2) | 0.015 |
| Ge | 71.1(2) | -153.8(2) | -2.3(2) | 0.019 |
| Sn | 76.2(2) | -153.0(2) | -0.3(2) | 0.020 |

As shown in **Table 1**, $J_1$ varies from negligible to strong FM when the X site changes from Si to Sn. Meanwhile, stronger AFM interaction $J_2$ remains unaffected by the X site, which reveals that the shape of the Gd orbitals changes primarily along the a or b axis when Si is replaced by Ge/Sn, suggesting an enhancement of $J_1$. Moreover, the color contrast along the edges (associated with $J_1$) and diagonals (associated with $J_2$) highlights differences in both sign and magnitude (**Figure 5b**). Along the body diagonal, weak AFM interaction $J_3$ shows minimal dependence on the X site (**Figure 5d**), as indicated by subtle color changes (small in magnitude) and similar patterns along the (111) direction. This observation aligns well with the magnetic properties of the Si and Ge compounds.

To assess the competing strength of these exchange interactions, we calculated the inverse of the Mean Absolute Deviation (MAD$^{-1}$), which measures the closeness of interaction strengths. As the X site progresses from Si to Ge and Sn, the MAD$^{-1}$ value improves from 0.015 to 0.019 and 0.020, respectively. This increase in the appreciable, competing FM and AFM interactions in the Gd square lattice (the *ab*-plane) as a function of X follows similar trends in enhanced electronic instability and destabilizing energy $E_{band}$. Competing FM and AFM interactions bear a resemblance to those in other skyrmion hosts.

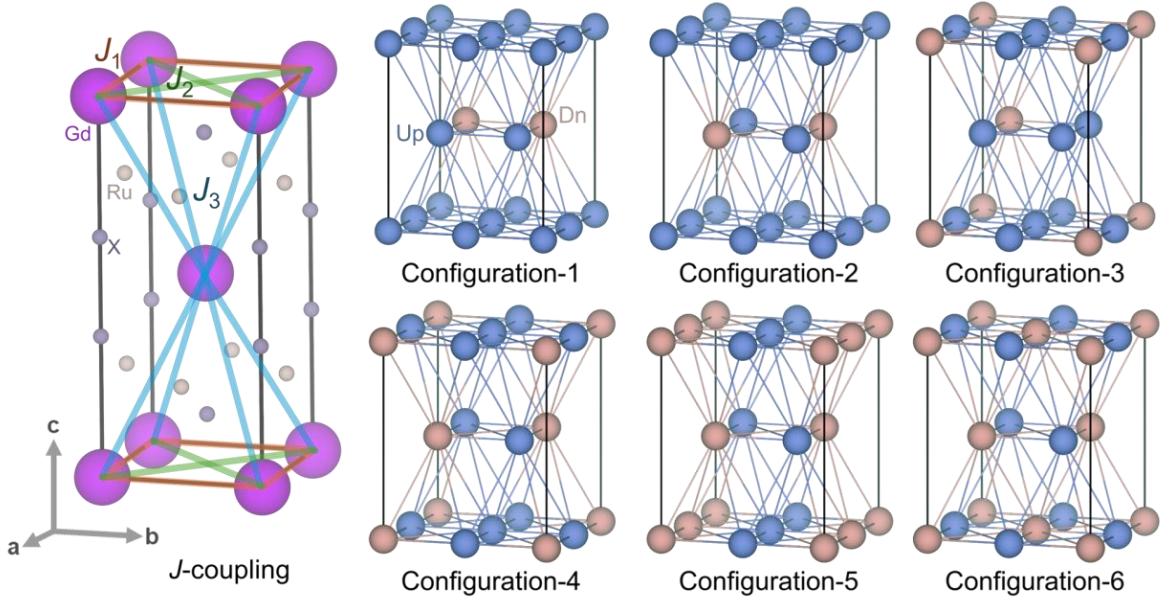

**Figure 10.** Representing exchange interactions $J_1$, $J_2$ and $J_3$ between $Gd^{3+}$ atoms within the unit cell and spin-ordered states within the (2a, 2b, c) supercell. Pink and blue colors correspond to the spin up and down, respectively.

We propose a correlation between $E_{band}$ and the temperature and magnetic field conditions at which skyrmions emerge (**Figure 11**). For the Si and Ge materials, the conditions of skyrmion formation were experimentally proven. Our attempts to make the Sn version were not fruitful yet, possibly due to its highest electron instability and thermodynamically unfavorable formation energy among the series. While we are making additional efforts to create $GdRu_2Sn_2$, we predict that this phase hosts skyrmions at a higher temperature and a lower field than its siblings, attributable to its highest destabilizing energy $E_{band}$. This prediction can be justified by the DOE formalism fundamentally painting the entire energetic picture.

Taken together, we propose a trend: the more destabilizing energy $E_{band}$ and the more competing FM and AFM interactions in the Gd square lattice, the more accessible skyrmion formation (higher temperature and lower field) (**Figure 11**). This suggestion demonstrates that destabilizing energy contributions and competing interaction strengths can serve as indicators for predicting and realizing the emergence of skyrmions. While the suggested trend may be proven useful, it warrants rigorous, experimental evidence for the Sn material.

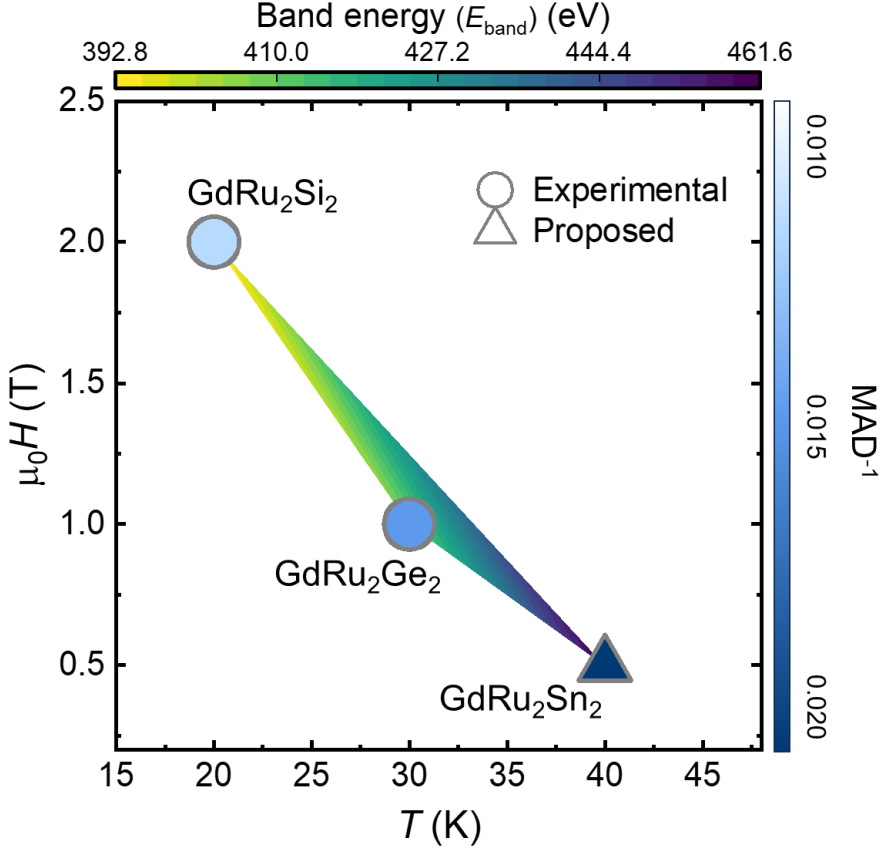

**Figure 11.** Correlation between skyrmion formation $T$, and $\mu_0 H$ with band energy ($E_{band}$) for GdRu$_2$X$_2$ (X = Si, Ge, and Sn), with the blue color filling indicate the MAD$^{-1}$.

**CONCLUSIONS**

Developing skyrmion materials for spintronics requires a comprehensive understanding of the chemical origins that facilitate a system to undergo a phase transition under a given condition to form the topologically distinct spin phases. Our results provide a new pathway toward understanding and developing centrosymmetric magnetic metals hosting skyrmions rooted in chemical concepts in solids. We study the impact of X-$p$ (Si-3$p$/Ge-4$p$/Sn-5$p$) orbitals on the skyrmion evolution in GdRu$_2$X$_2$. The chemical bonding analysis reveals that increased interactions between the Gd-4$f$ localized electrons and the [Ru$_2$X$_2$] conduction layer occur as X changes from Si-3$p$ to Ge-4$p$ and Sn-5$p$ (more extended orbitals). The realization of FSN proves the electron instability in GdRu$_2$X$_2$. In this analysis, an FSN vector [$Q = (q, 0, 0)$] and two inequivalent FSN vectors [$Q = (q, 0, 0)$; $Q_A (q, q, 0)$] are extracted for the Si and Ge compounds, respectively. For the Sn material, multiple FSN vectors are suggested. The competition between RKKY interactions at different wavevectors, stemming from FSN along several directions, can work in favor of skyrmion formation in centrosymmetric magnetic metals. The DOE analysis complements COHP in that it captures the significance of both interatomic and atomic energetic contributions for electron (in)stability. We propose that the more destabilizing energy and the more competing interaction strength, the more accessible conditions at which skyrmions emerge (at higher temperature and lower fields). Ongoing experimental studies on the Sn compound will help verify this proposed trend. Overall, our work provides a new framework to approach skyrmion materials research from chemical bonding and electronic instability perspectives while inviting other studies to dissect multifaceted aspects of the emergence of the topologically nontrivial states of matter for high-density memory and logic architectures.

## ASSOCIATED CONTENT

**Supporting Information**. Additional data analysis, tables, figures, including COBI, Fermi surface. This material is available free of charge via the Internet at http://pubs.acs.org.

## AUTHOR INFORMATION

**Corresponding Author**
*Thao T. Tran, email: thao@clemson.edu.**Corresponding Author**
*Thao T. Tran, email: thao@clemson.edu.

**Author Contributions**
The manuscript was written through the contributions of all authors. All authors have given approval to the final version of the manuscript.

## ACKNOWLEDGMENTS

The work at Clemson University was supported by the National Science Foundation under CAREER Award NSF-DMR-2338014. X.H. and R.K. thank the Arnold and Mabel Beckman Foundation for a 2023 BYI award to T.T.T. We acknowledge support from the Arnold and Mabel Beckman Foundation and the Camille and Henry Dreyfus Foundation. Research performed at Gdansk Tech was supported by the National Science Center (Poland) OPUS grant no. UMO-2022/45/B/ST5/03916. S.T. acknowledges support from ONR-N000142312061.

# Supporting Information for

# Antibonding and Electronic Instabilities in GdRu$_2$X$_2$ (X = Si, Ge, Sn): A New Pathway Toward Developing Centrosymmetric Skyrmion Materials

Dasuni N. Rathnaweera,[1] Xudong Huai,[1] K. Ramesh Kumar,[1] Sumanta Tewari,[2] Michał J. Winiarski,[3] Richard Dronskowski,[4] and Thao T. Tran[1],*

[1]Department of Chemistry, Clemson University, Clemson, South Carolina, 29634, United States.

[2]Department of Physics, Clemson University, Clemson, South Carolina, 29634, United States.

[3]Faculty of Applied Physics and Mathematics and Advanced Materials Center, Gdansk University of Technology, ul. Narutowicza 11/12, 80-233 Gdansk, Poland

[4]Chair of Solid-State and Quantum Chemistry, Institute of Inorganic Chemistry, RWTH Aachen University, 52056 Aachen, Germany

**Experimental section:**

List of Figures:

**Figure S1.** Formation energy of GdRu$_2$X$_2$ (X = Si, Ge, Sn).

**Figure S2.** Brillouin zone for body-centered tetragonal lattice.

**Figure S3.** Crystal orbital bond index (COBI) curves for a) GdRu$_2$Si$_2$, (b) GdRu$_2$Ge$_2$, and (c) GdRu$_2$Sn$_2$ with their integrated values (ICOBI).

**Figure S4.** Spin density map of (001) lattice plane.

**Figure S5.** Spin density map of (110) lattice plane.

**Figure S6.** Fermi surface of GdRu$_2$X$_2$. (a) spin up, (b) spin down band structure and corresponding fermi surfaces of GdRu$_2$Si$_2$. (c) spin up, (d) spin down band structure and corresponding fermi surfaces of GdRu$_2$Ge$_2$, and (e) spin up, (f) spin down band structure and corresponding fermi surfaces of GdRu$_2$Sn$_2$

**Figure S7.** Inequivalent Fermi surface nesting vectors of GdRu$_2$Ge$_2$ in band 4.

**Figure S8.** (a) and (b) Molecular orbital diagrams for GdRu$_2$Si$_2$, GdRu$_2$Ge$_2$ and GdRu$_2$Sn$_2$.

List of Tables:

**Table S1.** Crystal structure information of GdRu$_2$X$_2$ (X = Si, Ge, Sn).

**Density Functional theory calculations.**

Self-consistent spin-polarized electronic structure calculations were performed using a full-potential linearized augmented plane wave method by the WIEN2k code.[1] The exchange and correlation energies were treated within the density functional theory (DFT) using the Perdew–Burke–Ernzerhof generalized gradient approximation.[2] The self-consistencies were carried out using 9 × 9 × 4 $k$-mesh, for GdRu$_2$(Si/Ge)$_2$ and 8 × 8 × 3.5 for GdRu$_2$Sn$_2$ in the irreducible Brillouin zone. A Hubbard U correction of 6.7 eV was used for the strong correlations in 4$f$ bands. The muffin tin radius values were 1.32, 1.17, 0.96, 1.12, and 1.17 Å for Gd, Ru, Si, Ge, and Sn, respectively.

Pseudopotential band structure and polarized density of states were calculated using the pw.x program in the Quantum Espresso (QE) software package,[3] with the Generalized Gradient Approximation of the exchange-correlation potential and Hubbard U correction (GGA+U)for the strong correlations in 4$f$ bands (6.7 eV for Gd-4$f$)[4] of the exchange-correlation potential with the PBEsol parametrization.[5] Projector-augmented wave (PAW) potentials for Si, Ge, Sn and Ru were taken from the PSlibrary v.1.0.0 set, and for Gd, PAW potential developed by VLab was used.[6,7] Lattice parameters for GdRu$_2$Sn$_2$ were estimated using variable cell calculation. The applied $k$-mesh is the same as in WIEN2k calculations. Kinetic energy cutoff for charge density and wavefunctions was set to 30 eV and 360 eV. The pseudo potential DFT calculations for the J-coupling constant used the same parameter, except the k-mesh was changed to 3 × 3 × 2 to account for the (2a, 2b, c) super cell. The charge and spin density maps are calculated with the pp.x in QE package and visualized with VESTA software. The pseudopotential wavefunctions calculated from the QE package are projected into a linear combination of atomic orbitals (LCAO) based representation by means of Local Orbital Basis Suite Towards Electronic-Structure Reconstruction (LOBSTER)[8] program to extract the projected crystal orbital Hamilton populations (pCOHP)[9], crystal orbital bond index (COBI)[10], density of energy (DOE)[11] and the molecular orbital diagrams.

The Fermi surface was calculated using the fs.x software in the QE software package and visualized using XCrySDen software. Additional SCF calculations were performed to compute the Lindhard response function (LRF). The kinetic energy cutoff for the plane-wave basis was set to 40 Ry, while the cutoff for the charge density was set to 452 Ry. To extract the Fermi surface, the self-consistent field (SCF) calculation was performed using a denser Monkhorst-Pack (MP) k-point grid. The Brillouin zone was numerically integrated for the total energy estimation using a 38 × 38 × 19 MP k-point sampling. Fermi surface calculations were carried out based on the converged SCF results. The Fermi velocity across the Brillouin zone was mapped onto the Fermi surface as a color plot. The orbital character of the Fermi surface was computed using the Fermi projector operator, which calculates the following quantity: $\sum_{i=1}^{n_s} \left| \left\langle \emptyset_{n_{s(i)}}^{atom} \middle| \emptyset_{nk} \right\rangle \right|^2$ where $n_s$ represents the number of target wavefunctions. The results from this Fermi projection calculation are saved in a format compatible with the FermiSurfer 2.4.0 software for visualization.[12]

**Figure S1.** Formation energy of GdRu$_2$X$_2$ (X = Si, Ge, Sn).

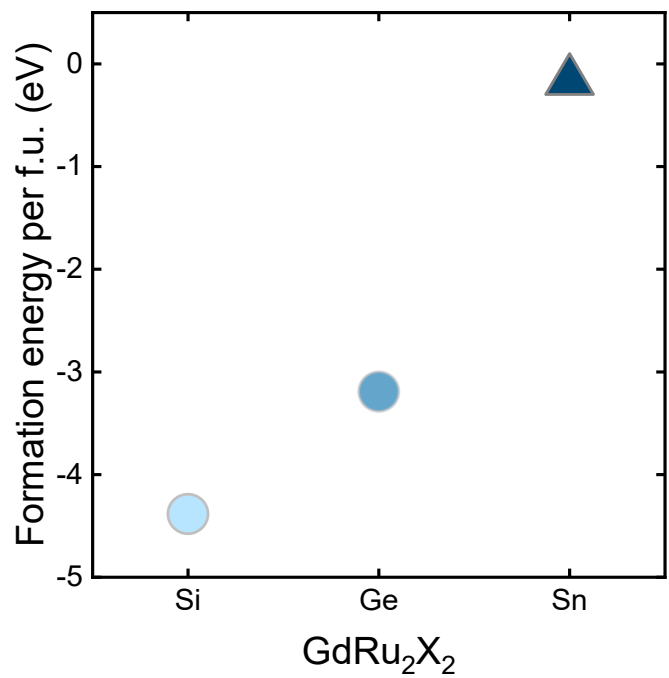

**Figure S2:** Brillouin zone for body-centered tetragonal lattice.[7, 13]

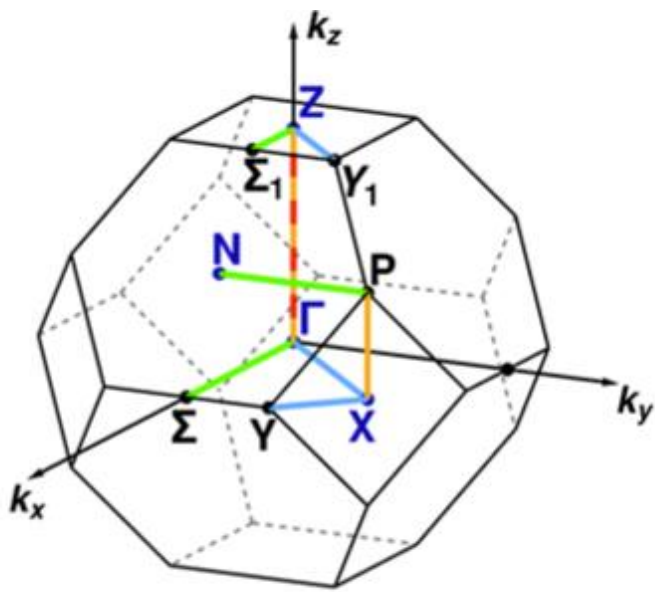

**Figure S3.** Crystal orbital bond index (COBI) curves for a) GdRu$_2$Si$_2$, (b) GdRu$_2$Ge$_2$, and (c) GdRu$_2$Sn$_2$ with their integrated values (ICOBI).

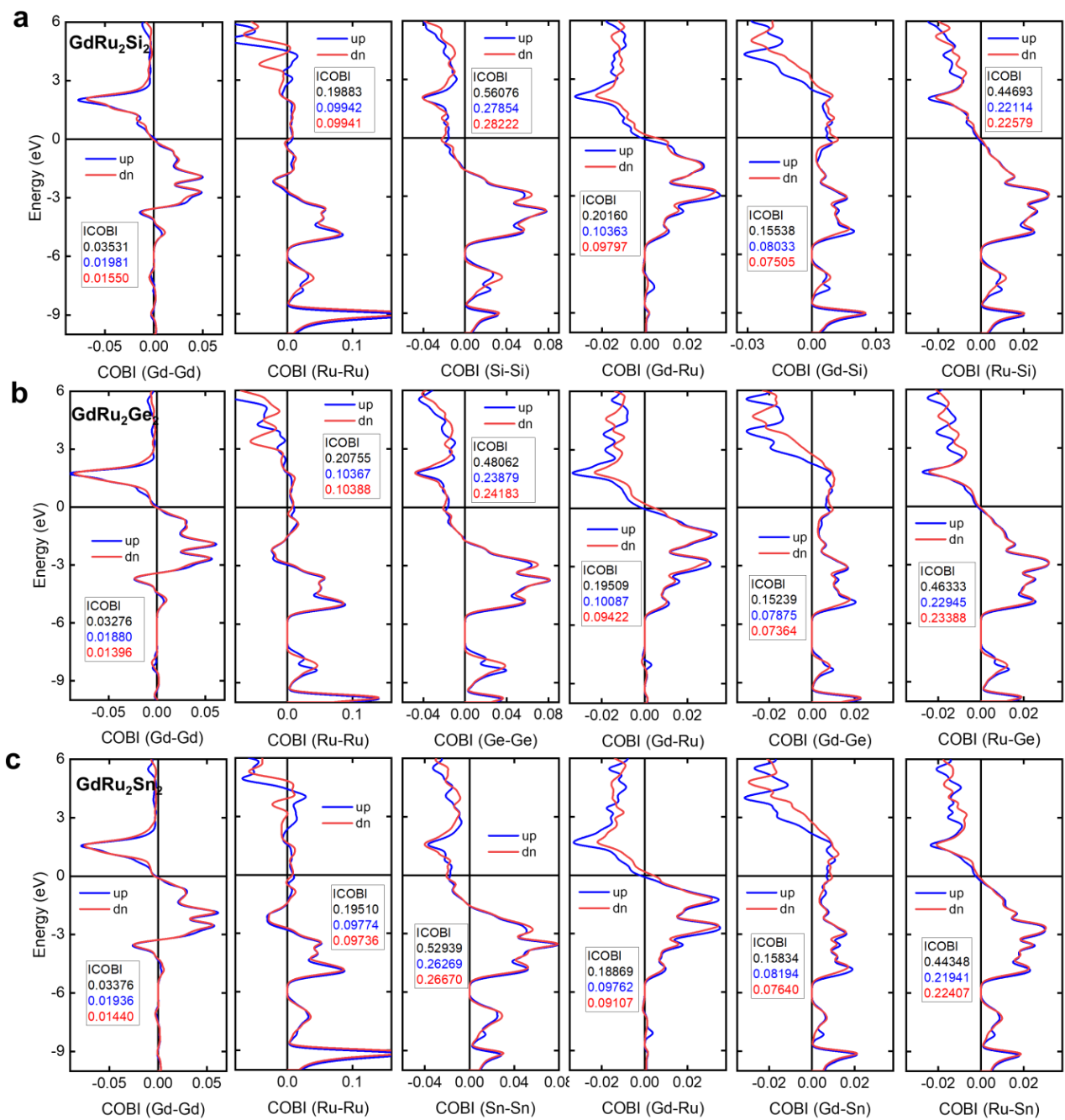

**Figure S4.** Spin density map of (001) lattice plane.

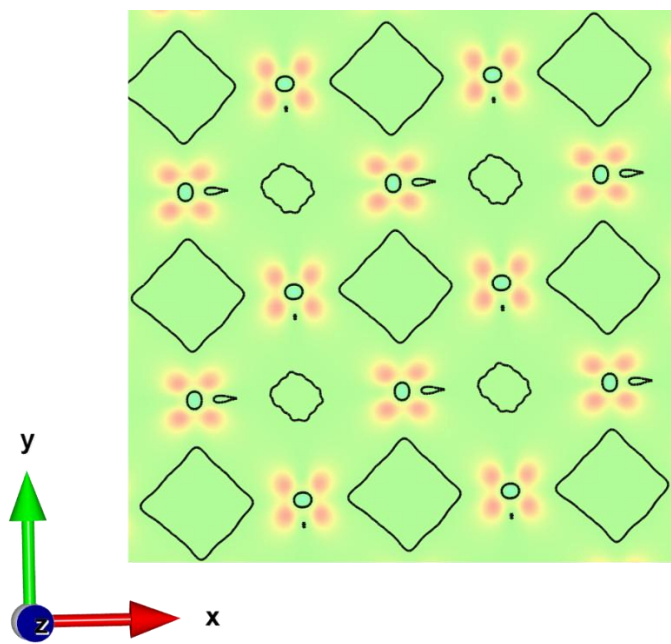

Orbital overlap in the [Ru$_2$X$_2$] layer was constructed based on this spin density map along (001)

**Figure S5.** Spin density map of (110) lattice plane.

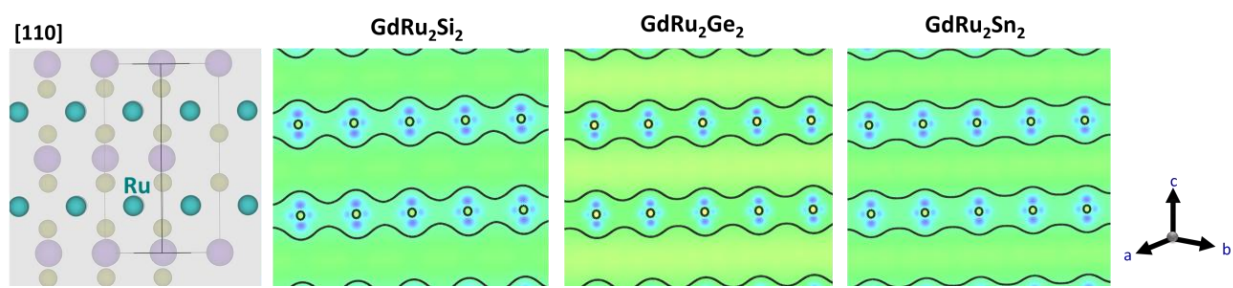

**Figure S6:** Fermi surface of GdRu$_2$X$_2$. (a) spin-up, (b) spin-down band structure and corresponding Fermi surfaces of GdRu$_2$Si$_2$. (c) spin-up, (d) spin-down band structure and corresponding Fermi surfaces of GdRu$_2$Ge$_2$, and (e) spin-up, (f) spin-down band structure and corresponding Fermi surfaces of GdRu$_2$Sn$_2$

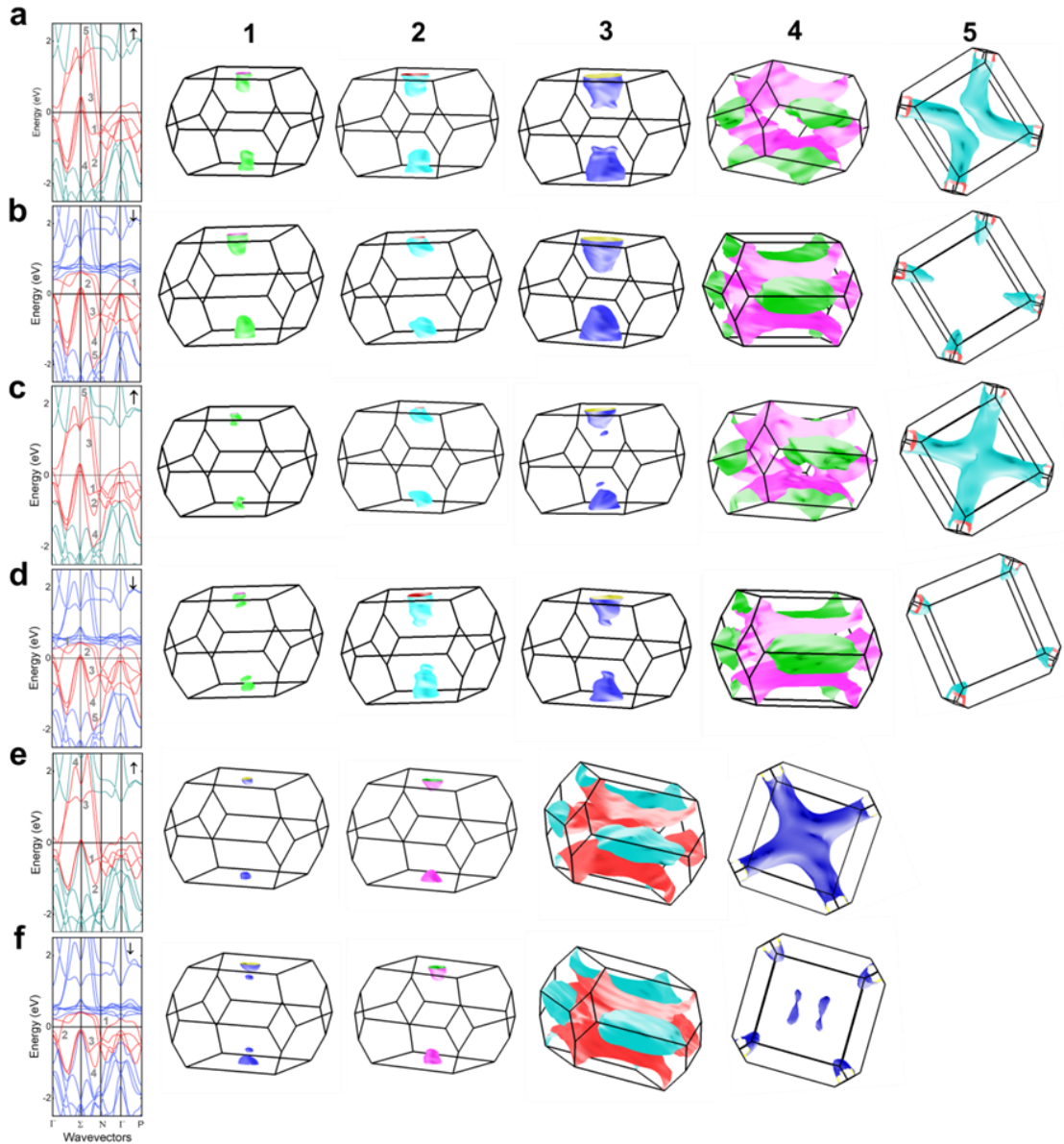

**Figure S7.** Inequivalent Fermi surface nesting vectors of GdRu$_2$Ge$_2$ in band 4.

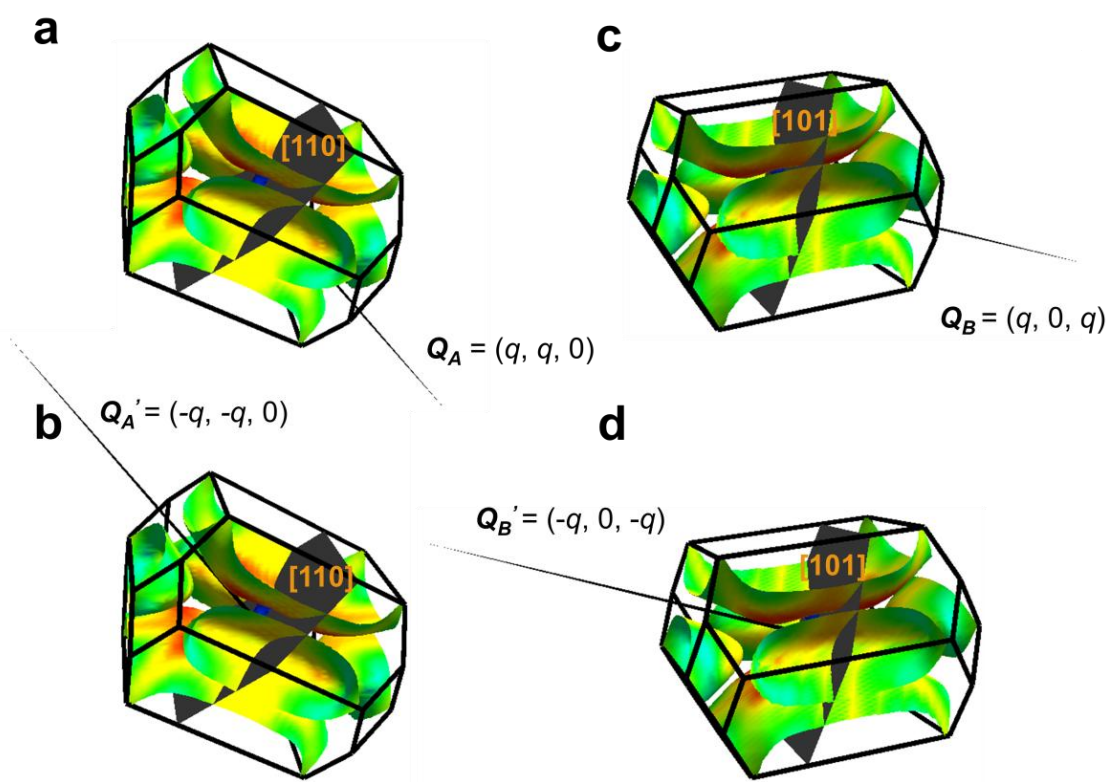

**Figure S8.** (a) and (b) Molecular orbital diagrams for GdRu$_2$Si$_2$, GdRu$_2$Ge$_2$ and GdRu$_2$Sn$_2$.

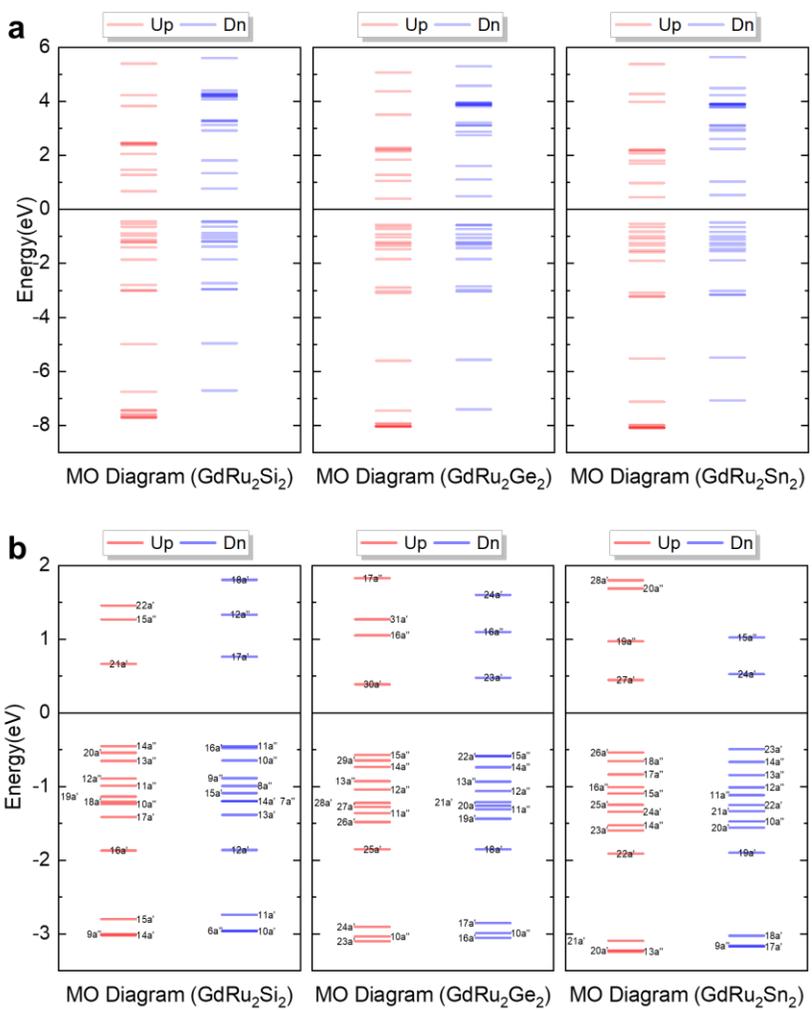

**Table S1.** Crystal structure information of GdRu$_2$X$_2$ (X = Si, Ge, Sn).

| Empirical formula | GdRu$_2$Si$_2$* | GdRu$_2$Ge$_2$ | GdRu$_2$Sn$_2$** |
|---|---|---|---|
| Formula weight | 415.56 | 504.608 | 596.81 |
| Crystal system | Tetragonal | Tetragonal | Tetragonal |
| Space group | *I*4/*mmm* | *I*4/*mmm* | *I*4/*mmm* |
| $a$/Å | 4.164(2) | 4.23204(12) | 4.31147 |
| $c$/Å | 9.616(5) | 9.8643(4) | 10.02388 |
| $V$/Å$^3$ | 166.73 | 176.671(10) | 186.33 |
| Z | 2 | 2 | 2 |
| $\rho_{calc}$ /g cm$^{-3}$ | 8.308 | 9.486 | 10.48049 |

* database_code_ICSD 636306

** Based on variable cell calculations